# The low-energy theory for the Bose-Hubbard model and the normal ground state of bosons

Igor F. Herbut
Department of Physics, Simon Fraser University, Burnaby, British Columbia,
Canada V5A 1S6

**Abstract**: A bosonic realization of the $SU(2)$ algebra and of its vector representation is constructed, and an effective low-energy description of the Bose-Hubbard model in the form of anisotropic theory of quantum rotors is proposed and discussed. A possibility of a normal zero-temperature bosonic phase with neither crystalline nor superfluid order around the tip of the checkerboard-solid lobe at half-integer fillings is examined.

The ground state of interacting bosons is typically a superfluid, unless a sufficiently long-ranged interparticle interaction or a commensurate periodic potential is present, in which case bosons can form a crystalline solid even at $T = 0$. The simplest description of the interacting bosonic systems is provided by the Bose-Hubbard model and its variants, which depending on range and strengths of interactions and on density of particles exhibit uniform Mott insulating, charge-density-wave (solid), superfluid, and mixed supersolid ground states [1]. Influence of quantum fluctuations on the phase diagram of quantum bosons has been somewhat less well understood, and recently a possibility of a normal phase of bosons at $T = 0$ has been raised in the context of experiments on superconductor-insulator transition in thin films [2]. The normal phase would be without either solid or superfluid order, and, presumably, would show the conducting properties of a metal. Leggett has supplied arguments against such a ground state for bosons in the continuum some time ago [3], but on a lattice the situation is less clear. The issue is not only of theoretical interest, as there exist many experimental realizations of lattice bosonic systems, including granular superconductors [4], Josephson junction arrays [5], or, since recently, cold atoms in optical lattices [6]. Recent numerical calculations which focused on this question [7], [8] in $D = 1$ gave results that seem to depend on the details of the model, and no coherent picture has yet emerged.

It is well known, on the other hand, that for infinite on-site repulsion, which limits the boson occupation numbers at each site to zero and one, bosonic system becomes equivalent to the system of spins-1/2. Das and Doniach [2] argued that in the opposite limit of a large number of particles per site, the equivalent quantum-phase (or Josephson junction array) model can be written in terms of the generators of the Euclidean group $E_2$, which again is intimately related to, in fact just a simple contraction of, the group $SO(3)$. It appears natural to ask if there exists a more general connection between the rotational symmetry and the lattice bosons. One such connection is established in this paper, where I display a bosonic realization of the $SU(2)$ Lie algebra and show that the two-component superfluid order parameter and the order parameter for the checkerboard solid phase can be combined into this algebra's three dimensional vector representation. Similar operators were constructed previously for the electronic Hubbard model [9], and applied to its negative-U version. Here, it is proposed that near half-integer fillings the Bose-Hubbard model with on-site and nearest-neighbor repulsion may posses an approximate $SU(2)$ symmetry, and that the effective low-energy theory near these points has the form of anisotropic Hamiltonian for quantum rotors. At half-integer fillings, where the mean-field theory predicts a first-order (spin-flop) transition between solid and superfluid phases, the effect of quantum fluctuations is the strongest, and right at the boundary it can produce a disordered phase, with only short-range crystalline or superfluid order. Since such a phase has a finite gap and therefore is stable to small perturbations, it will occupy a finite region of the phase diagram. In this case the direct solid-to-superfluid transition is replaced by two continuous (of the Ising and $XY$ variety) transitions between solid and normal, and normal and superfluid phases. In the example of hard-core bosons with nearest-neighbor and weak next-nearest-neighbor interactions, the effective rotor theory suggests that the intermediate normal phase is most likely absent in dimensions $D > 1$. I speculate that in $D = 1$ the normal phase may be observable at larger filling factors or for weaker on-site repulsion.

Consider a system of bosons on a quadratic lattice in $D$-dimensions and at $T = 0$. It is straightforward to check that the operators:

$$L_+ = \frac{i}{2} \sum_{\vec{p}} b^\dagger_{-\vec{p}+\vec{Q}} b^\dagger_{\vec{p}}, \tag{1}$$

$$L_- = \frac{i}{2} \sum_{\vec{p}} b_{\vec{p}} b_{-\vec{p}+\vec{Q}}, \tag{2}$$

$$L_3 = \frac{1}{2}(\sum_{\vec{p}} b^\dagger_{\vec{p}} b_{\vec{p}} + \frac{N_L}{2}), \tag{3}$$

form the $SU(2)$ Lie algebra: $[L_3, L_\pm] = \pm L_\pm$, $[L_+, L_-] = 2L_3$. Here $[b_i, b_j^\dagger] = \delta_{ij}$ are the standard Bose operators, $b_i = \sum_{\vec{p}} \exp(i\vec{p} \cdot \vec{R}_i) b_{\vec{p}}$, $\vec{Q} = (\pi/a, \pi/a....\pi/a)$ and the sums over momenta run over the first Brillouin zone. $N_L$ is the total number of lattice sites. Components of the three dimensional vector representation of $SU(2)$ should satisfy the commutation relations: $[L_\alpha, n_\beta] = i\epsilon_{\alpha\beta\gamma} n_\gamma$, where $\alpha = 1, 2, 3$, and $L_\pm = L_1 \pm iL_2$. A possible $SU(2)$ vector multiplet is then:

$$n_1 = \frac{1}{2} \sum_i (b_i^\dagger b_i^\dagger + b_i b_i), \tag{4}$$

$$n_2 = \frac{1}{2i} \sum_i (b_i^\dagger b_i^\dagger - b_i b_i), \tag{5}$$

$$n_3 = \frac{1}{i} \sum_i e^{i\vec{Q} \cdot \vec{R}_i} b_i^\dagger b_i. \tag{6}$$

The third component is proportional to the amplitude of the charge-density-wave with the period of two lattice spacing in every direction, and represents the order parameter for the checkerboard-solid phase. It is interesting that in order to satisfy the $SU(2)$ commutation relations it is necessary for $n_1$ and $n_2$ to be *bilinear* in Bose operators. The finite expectation value of $n_1$ or $n_2$ thus signals the long-range order in the composite operator $bb$, not $b$ itself. If there is no attractive interaction in the microscopic Hamiltonian that could lead to pairing of bosons it seems plausible to assume that the composite operator $bb$ can condense *only* when $\langle b \rangle \neq 0$. I will take this to be the case and hereafter make no difference between the two forms of superfluid order.

In general, the Bose-Hubbard Hamiltonian

$$H = -t \sum_{\langle i,j \rangle} b_i^\dagger b_j + \frac{1}{2} \sum_{i,j} v_{ij} n_i n_j - \mu \sum_i n_i, \tag{7}$$

of course commutes with $L_3$, but not with $L_1$ and $L_2$, and does not posses the exact $SU(2)$ symmetry. However, the numerical calculations and the mean-field theory [1] show that right at half-integer fillings the typical system described by the Bose-Hubbard model exhibits solid and superfluid phases at $T = 0$, which brake Ising and $U(1)$ symmetry, respectively. The discrete $Z_2$ and the $U(1)$ symmetries can be embedded into a larger $SU(2)$ group, and the existence of the bosonic version of the SU(2) algebra and of its vector multiplet allows one to construct a phenomenological theory for the Bose-Hubbard model that would apply at low energies and provide a unified description of solid and superfluid order. In continuum notation, the effective Hamiltonian density assumes the form:

$$H_{eff} = \frac{1}{2} E_{cl}[\vec{n}] + \frac{1}{2}(\frac{1}{\chi_\alpha} L_\alpha^2 + \rho_\alpha (\nabla_i n_\alpha)^2) - h_\alpha L_\alpha, \tag{8}$$

where $\vec{n}$ and $\vec{L}$ are dependent on the $D$-dimensional coordinate $\vec{x}$ and the imaginary time $\tau$, and the Hamiltonian is obtained by integration over space and imaginary time, as usual. $\vec{L}$ can be thought of as the angular momentum. We allow that the $SU(2)$ symmetry is broken to $U(1) \times Z_2$ both implicitly, by $\chi_{1,2} = \chi_\perp \neq \chi_3$ and $\rho_{1,2} = \rho_\perp \neq \rho_3$, and explicitly, by including the source fields $\vec{h}$ and the classical energy $E_{cl}$, which favor a particular orientation of the vector $\vec{n}$. I assume the simplest form $E_{cl} = g n_3^2$. Also, the fluctuations of the order parameter amplitude will be taken to represent high-energy modes, and only the orientation of $\vec{n}$ then remains at low energies. Thus $\vec{n}^2 = 1$, and $\vec{L} \cdot \vec{n} = 0$.

Since the rotor Hamiltonian (8) is quadratic in $\vec{L}$ one may integrate the generators out to obtain the effective theory for the order parameters only. This requires some care because of the orthogonality between $\vec{L}$ and $\vec{n}$. I introduce a "gauge-field" as $\vec{L} = \vec{n} \times \vec{A}$, and add a "gauge-fixing" term to the $H_{eff}$ as $(1/2\omega)(\vec{n} \cdot \vec{A})^2$. The integration over $\vec{A}$ can then be performed, and it results in the Lagrangian density

$$L_{eff} = \frac{1}{2} E_{cl}[\vec{n}] + \frac{1}{2}(\rho_\alpha (\nabla_i n_\alpha)^2 + \chi_\perp (\dot{\vec{n}} - i\vec{n} \times \vec{h})^2 + \tag{9}$$

$$\chi_\perp \frac{v}{1 - v n_\perp^2} (\vec{n} \times \dot{\vec{n}} - i \vec{n} \times \vec{n} \times \vec{h})_3^2),$$

with $v = 1 - (\chi_\perp/\chi_3)$, independent of the gauge-fixing parameter $\omega$. Similar form has been studied by Zhang in context of his $SO(5)$ theory of high-$T_c$ superconductivity [10], but even in the classical approximation $\dot{\vec{n}} = \nabla \vec{n} = 0$ the Eq. 9 differs from his expression [11]. Only if the symmetry breaking is weak, i. e. $v << 1$, the classical ground states of the system are determined by minimization of a simpler expression equivalent to Zhang's:

$$L_{cl} \approx \frac{1}{2} g n_3^2 - \frac{h^2}{2} n_\perp^2 (\tilde{\chi}_\perp n_3^2 + \tilde{\chi}_3 n_\perp^2), \tag{10}$$

where $\tilde{\chi}_\perp = \chi_\perp$, $\tilde{\chi}_3 = \chi_\perp(1 + v)$, and $\vec{h} = (0, 0, h)$. At $h = 0$ there is a first-order (spin-flop) transition from solid to superfluid as $g$ changes sign from negative to positive. At finite $h$ and if $\tilde{\chi}_3 > \tilde{\chi}_\perp$, the same remains true only at negative $g$, while if $\tilde{\chi}_3 < \tilde{\chi}_\perp$ there exists an intermediate supersolid phase in between.

The effect of quantum fluctuations on the classical solutions is expected to be the strongest right at the point $g = h = 0$, where the classical energies of solid and superfluid phases become degenerate. After rescaling the imaginary time as $(\rho_\perp/\chi_\perp)^{1/2} \tau \to \tau$ the fluctuation part of the Lagrangian becomes:
$$L_{eff} = \frac{1}{2T_{eff}}((\nabla_i n_\alpha)^2 + (\dot{\vec{n}})^2 + u(\nabla_i n_3)^2 + v(\vec{n} \times \dot{\vec{n}})_3^2), \tag{11}$$

where $u = (\rho_3/\rho_\perp) - 1$. It is the non-linear sigma model in $D+1$-dimensions, at an effective temperature $T_{eff} = 1/(\rho_\perp \chi_\perp)^{1/2}$, with two new terms representing the remaining implicit symmetry breaking. To determine the phase of the system at the point $h = 0$ I perform the renormalization group transformation on the model, near $D = 1$ and perturbatively in $T_{eff}$, $u$ and $v$ [12]. The integration over the degrees of freedom within the momentum shell $(\Lambda/s, \Lambda)$ changes the couplings of the model according to the differential equations:

$$\frac{dT_{eff}}{d\ln(s)} = (1-D)T_{eff} + \frac{m-2+\theta(g)(m-3)g}{2\pi(1+|g|)}T_{eff}^2, \tag{12}$$

$$\frac{d\{u,v\}}{d\ln(s)} = -\frac{1}{4\pi}\{3u, (m-2)v\}T_{eff}, \tag{13}$$

$$\frac{d|g|}{d\ln(s)} = 2|g| - \frac{T_{eff}|g|}{\pi(1+|g|)}, \tag{14}$$

where the general number of components $m$ is retained, $m = 3$ being the case in point. $\theta(g)$ is the step function, and I redefined $gT_{eff}/\Lambda^2 \to g$. Weak implicit symmetry breaking is, expectedly [13], [14], irrelevant, and in $D > 1$ at $g = 0$ the system has a phase transition at the critical effective temperature $T_{eff}^* = 2\pi(D-1) + O((D-1)^2)$. For $T_{eff} < T_{eff}^*$ the fluctuations are too weak to change the mean-field result, and there is indeed a direct first-order transition between solid and superfluid at $g = 0$. In the opposite case quantum fluctuations qualitatively change the phase diagram. Since the disordered phase, which then exists right at $g = h = 0$, has a finite correlation length and thus a finite gap, it can not collapse at infinitesimal $g$ or $h$. Consequently, the disordered phase spreads from the point $g = h = 0$ to a finite region around it, and gives way to the solid or the superfluid only at some finite values of $g$ or $h$. This also becomes evident from the above recursion relations at finite $g$, which reduce to those in ref. 13 when $g > 0$. The resulting flow diagram for $h = u = v = 0$ is depicted for $m = 3$ and $D > 1$ at Figure 1. By increasing $g$ at fixed $T_{eff} > T_{eff}^*$ one intersects both solid-to-normal and normal-to-superfluid phase boundaries, and the normal phase persists for a finite interval of $g$.

Whether or not will a particular example of the Bose-Hubbard model exhibit the intermediate normal phase depends on the effective temperature at the symmetric point $g = h = 0$. In general, there exists no simple way to calculate the effective coupling constants in terms of the microscopic interactions. For the hard-core bosons with infinite on-site repulsion, however, the connection between microscopic and effective couplings can be made more explicit, due to the well known analogy with the spin-1/2 system. Taking $v_{ij} = V$ for $(i,j)$ nearest neighbors, and $v_{ij} = V' << V$ for the next-nearest neighbors, one can derive the effective rotor Hamiltonian in Eq. 8 in large spin-S approximation [15], and eventually take $S = 1/2$. This way one finds:

$$E_{cl} = -S^2(2ztn_\perp^2 + (zV - z'V')n_3^2), \tag{15}$$

$$h = \mu - \frac{1}{2}(zV - z'V'), \qquad (16)$$

$$\chi_\perp^{-1} = 2zt - S^{-2}E_{cl}, \qquad (17)$$

$$\chi_3^{-1} = zV + z'V' - S^{-2}E_{cl}, \qquad (18)$$

$$\rho_\perp = S^2 t, \qquad (19)$$

$$\rho_3 = \frac{S^2}{2}(V - V'). \qquad (20)$$

$z$ and $z'$ are the number of nearest and next-nearest neighbors on the lattice, respectively. For $V' = 0$, inserting these expressions into the Eq. 9, the classical approximation yields the first-order transition from solid to superfluid at:

$$t_c = \frac{1}{2}\sqrt{\mu(V - \frac{\mu}{2})} \qquad (21)$$

for $S = 1/2$, in *exact* agreement with the result of the mean-field approximation applied directly to the Bose-Hubbard Hamiltonian, in two dimensions [1]. In this case at half-filling (at $\mu = 2V$) and at $t = V/2$ the system has the full rotational symmetry, and in its immediate vicinity $\tilde{\chi}_3 = \tilde{\chi}_\perp$, so that there is no supersolid phase. When small $V' > 0$ is turned on, exactly at half-filling (now at $\mu = (zV - z'V')/2$) the solid-to-superfluid transition at $2t = V - z'V'/z$ remains first-order, but since $\tilde{\chi}_3 = \tilde{\chi}_\perp(1 - z'V'/zt) < \tilde{\chi}_\perp$, away from half-filling there immediately appears a thin sliver of the supersolid phase in between. Again, this is precisely what the direct mean-field treatment of the Bose-Hubbard model in the hard-core limit obtains, and our effective theory provides a very simple explanation for the known mean-field phase diagram.

Given the good agreement at the mean-field level we may attempt to estimate the effective temperature for hard-core bosons at the transition point at half-filling. Near $D = 1$ where the effective temperature is dimensionless, for $S = 1/2$ I find $T_{eff} = 2\sqrt{2z}$, which suggests that in $D > 1$ hard-core bosons with weak next-nearest-neighbor repulsion are probably too "cold" to have the intermediate disordered phase. For example, in $D = 2$ $T_{eff} \approx 4\sqrt{2}$ while $T_{eff}^* \approx 2\pi$, according to this very crude estimate. The case of $D = 1$ is special, since the presence of additional topological terms in the effective action makes the disordered phase at the solid-superfluid boundary gapless for half-integer spins [15]. Instead, in $D = 1$ one finds the continuous Kosterlitz-Thouless solid-superfluid transition right at half-filling [1], [8]. If the constraint on double occupancy is relaxed the Bose system is no longer exactly equivalent to the spin-1/2 Heisenberg Hamiltonian, but one would expect that the form of the low-energy theory should remain the same. For smaller on-site repulsion and larger half-integer fillings the multiple occupancy of the same site should become frequent, and it is conceivable that the rotor Hamiltonian then may become appropriate even in $D = 1$. This would suggest that the under these conditions the intermediate normal phase may appear at the tip of the solid-superfluid lobe, in qualitative agreement with the Monte Carlo calculations on the quantum-phase model of ref. 7 [16].

The normal phase of bosons discussed here differs in detail from the one proposed by Das and Doniach [2], which appears at integer filling in the quantum phase model for large enough nearest-neighbor repulsion. Both phases are the results of quantum disordering of the bosonic ground state, however, and should show similar physical properties. Studies of the conducting properties of this phase and of the effects of disorder [17] are left for future research.

In conclusion, I presented a bosonic representation of the $SU(2)$ Lie algebra, and showed that the superfluid and solid order parameters together form its vector representation. It was argued that the low-energy effective theory of the Bose-Hubbard model near half-integer fillings has the form of anisotropic Hamiltonian for quantum rotors. In classical approximation, the phase diagram derived from the effective theory agrees well with the previous studies. A possibility of a quantum disordered, or a normal, phase between solid and superfluid is pointed out, and the conditions for its appearance are briefly discussed.

This work has been supported by NSERC of Canada.

FIGURE CAPTIONS:

Figure 1: Schematic flow and phase diagram of the effective rotor theory at half-integer filling $h = 0$. At high effective temperature $T_{eff}$, with tuning of the parameter $g$ the system undergoes the solid-to-normal phase transition governed by the Ising critical point in $D+1$, followed by the normal-to-superfluid transition in the XY universality class. At low $T_{eff}$ the trivial fixed point governs the direct first-order solid-to-superfluid transition.

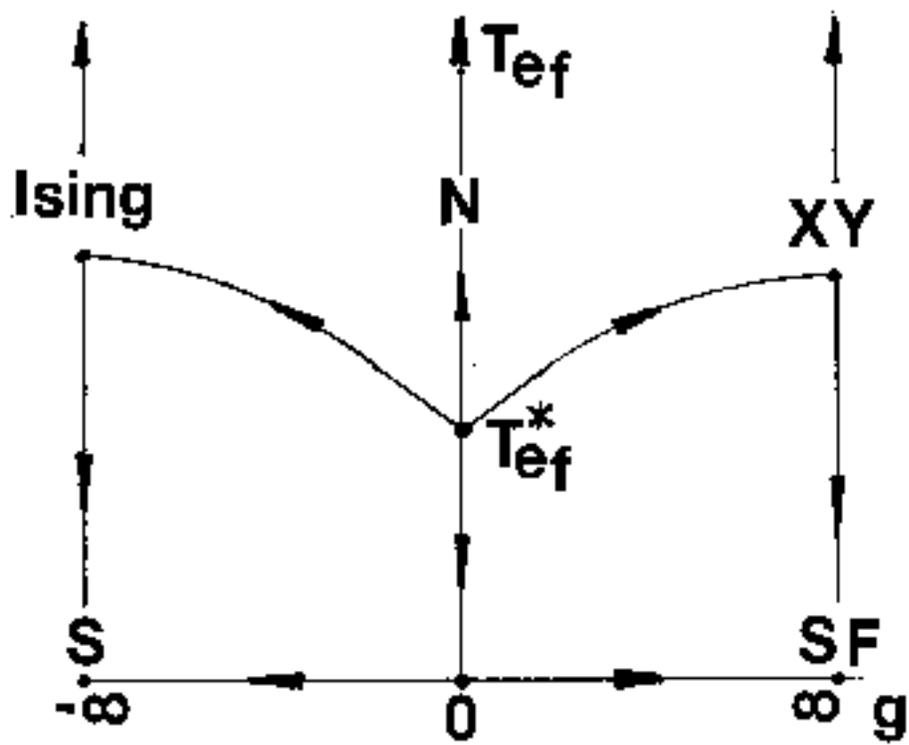